# Order from a mess: the growth of 5-armchair graphene nanoribbons

Alejandro Berdonces-Layunta,[1,2,†] Fabian Schulz,[3,4,†,*] Fernando Aguilar-Galindo,[1,†] James Lawrence,[1,2] Mohammed S. G. Mohammed,[1,2] Matthias Muntwiler,[5] Jorge Lobo-Checa,[6,7] Peter Liljeroth,[3] Dimas G. de Oteyza.[1,2,8,*]

[1] Donostia International Physics Center, 20018 San Sebastián, Spain

[2] Centro de Física de Materiales, 20018 San Sebastián, Spain

[3] Department of Applied Physics, Aalto University, FI-00076 Aalto, Finland

[4] Fritz Haber Institute of the Max Planck Society, 14195 Berlin, Germany

[5] Paul Scherrer Institute, 5232 Villigen, Switzerland

[6] Instituto de Nanociencia y Materiales de Aragón, 50009 Zaragoza, Spain

[7] Dpto. de Física de la Materia Condensada, Universidad de Zaragoza, 50009 Zaragoza, Spain

[8] Ikerbasque, Basque Foundation for Science, Bilbao, Spain

† these authors contributed equally

* correspondence to: schulz@fhi-berlin.mpg.de, d_g_oteyza@ehu.es

**Abstract:**

The advent of on-surface chemistry under vacuum has vastly increased our capabilities to synthesize carbon-nanomaterials with atomic precision. Among the types of target structures that have been synthesized by these means, graphene nanoribbons (GNRs) have probably attracted the most attention. In this context, the vast majority of GNRs have been synthesized from the same chemical reaction: Ullmann coupling followed by cyclodehydrogenation. Here, we provide a detailed study of the growth process of 5-atom-wide armchair GNRs starting from dibromoperylene. Combining scanning probe microscopy with temperature-dependent XPS measurements and theoretical calculations, we show that the GNR growth departs from the conventional reaction scenario. Instead, precursor molecules couple by means of a concerted mechanism whereby two covalent bonds are formed simultaneously, along with a concomitant dehydrogenation. Indeed, this novel reaction path is responsible for the straight GNR growth, something remarkable considering the initial mixture of reactant isomers with irregular metal-organic intermediates that we find. The provided insight will not only help understanding the reaction mechanisms of other reactants, but also serve as a guide for the design of new precursor molecules.

Carbon-based nanostructures have attracted enormous attention over the last few years due to the remarkable tunability of their properties and their associated potential for applications in modern technologies. The realization of diverse one-dimensional graphene nanoribbons (GNR) has been particularly successful,[1–3] including benzenoid structures with different edge orientations,[4–6] widths,[7,8] heteroatoms[9,10] and functional groups,[11,12] as well as other non-benzenoid $sp^2$-hybridized carbon backbones.[13–15] Each of these variations leads to materials with distinct optoelectronic and even magnetic properties.[1–3] Focusing on benzenoid nanoribbons with armchair-oriented edges, the lateral confinement of charge carriers causes the opening of a bandgap in graphene's originally quasi-metallic band structure,[16,17] which in turn allows for their application in field-effect transistors.[18,19] The bandgap is strongly width-dependent,[16,17,20] and 5-atom-wide armchair GNRs (5-AGNR) are particularly promising for such applications due to their low bandgap value.[16,17,21]

When the GNR synthesis is performed under vacuum, the most extensively used protocol employs the Ullmann-type coupling mechanism followed by cyclodehydrogenation (CDH) reactions on the surface of noble metal substrates, most commonly Au(111).[1–3] The former consists of a metal-catalyzed polymerization reaction guided by halogenated substituents in the reactant.[22,23] As the surface is annealed, a homolytic cleavage of the carbon-halogen bond is triggered, catalyzed by the metallic substrate. With few exceptions,[24] in the reported Ullmann-coupling experiments on Au(111) the halogen abstraction is directly followed by the formation of C-C bonds without a measurable intermediate. The subsequent CDH step normally displays a larger activation barrier, which requires substantially higher temperatures.[25–27]

In this work we study the growth process in the synthesis of 5-AGNRs and find out relevant differences as compared to other nanoribbons. To date, two different types of reactants have been used for the synthesis of 5-AGNRs.[28,29] The first one is 1,4,5,8-tetrabromonaphthalene (a tetrabrominated precursor) that does not require the CDH step to form the final product.[28] Instead, it displays an uncommon intermediate on Au(111), namely an organometallic oligomer that forms after dehalogenation and converts into 5-AGNRs in a second reaction step at higher temperatures.[28] The second reactant is dibromoperylene (DBP) halogenated at the 3 and 9 or 3 and 10 positions (Fig. 1a,b), which we use and study throughout this work. The sublimation powder employed for the GNR synthesis consisted in a mixture of these two DBP isomers[29]. In line with most other GNR precursor molecules explored to date, this reactant requires CDH processes for the ultimate ribbon formation. However, we find here that the growth process does not follow the conventional reaction pathway consisting of halogen activation, polymerization and subsequent CDH. Contrarily, by means of bond-resolving scanning probe microscopy (SPM) techniques and temperature-dependent XPS, we demonstrate that the DBP reaction on Au(111) starts with the formation of a metal-organic (MO) intermediate (Fig. 1c,d). Thereafter, a concerted mechanism sets in that leads to a double C-C bond formation involving a concomitant CDH reaction. This mechanism translates into surprisingly regular graphene nanoribbons despite the atomically inhomogeneous side terminations of these precursor (Fig. 1e). Particularly, three phases in the ribbon's growth dynamics are identified, starting with a seeding phase fed by the progressive dehalogenation of the precursor, and terminating by a fast growth in length due to the longitudinal fusion of the oligomers with still active ends.

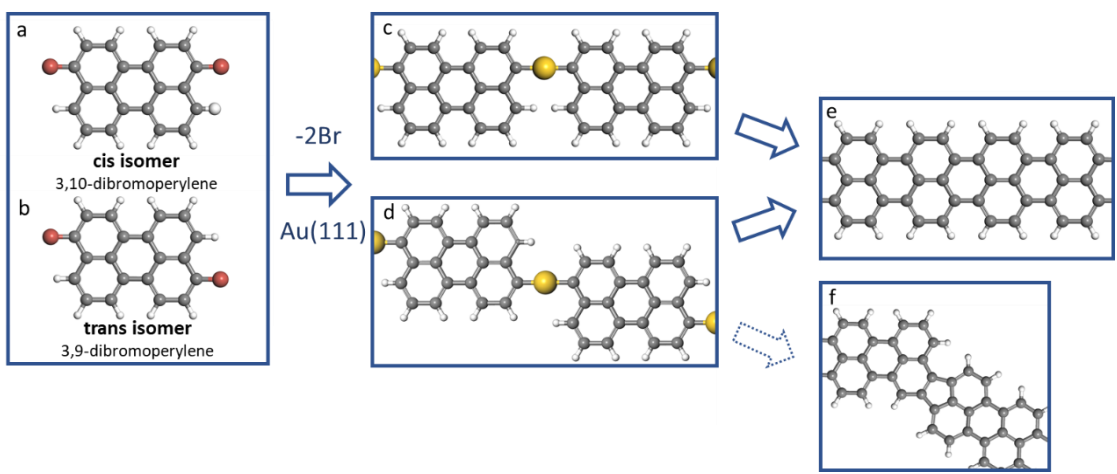

**Fig. 1**: *Illustrative scheme of the steps in the 5-aGNR formation pathway. (a,b) Carbon-halogen bonds of 3,9- or 3,10-dibromoperylene are cleaved when the molecule is annealed on Au(111). (c,d) The resulting diradical is stabilized by forming metal-organic chains. (e,f) In the same step, graphene nanoribbons are also formed, which suggests that dehydrogenation/C-C addition occurs simultaneously or is mediated by an unstable short-lived intermediate.*

# Results

For the SPM experiments, we deposited sub-monolayer coverages of DBP onto a clean Au(111) surface held at 250 K, followed by a stepwise annealing to increasing temperatures. After each annealing step, the sample was cooled down to 5 K for characterization. We used non-contact atomic force microscopy (AFM) with CO-functionalized tips for resolving the molecular backbone of the GNRs.[30]

Figure 2a shows a constant-current scanning tunneling microscopy (STM) image of the as-deposited reactants. The molecules self-assemble into porous networks with short-range order. Typical motifs are formed by clusters of three, four or six molecules. As can be seen in the constant-height AFM images in Figs. 2b and 2c, the bromine atoms – corresponding to the bright protrusions in the AFM images - are still attached to the molecule and stabilize the self-assembled structures *via* halogen bonds. Interestingly, we find that *cis* isomers (Fig. 1a) tend to form four-membered cavities (Fig. 2b) and *trans* isomers (Fig. 1b) six-membered ones (Fig. 2c). We ascribe this behavior to steric hindrance between neighboring DBP molecules. Cavities formed by a combination of *cis* and *trans* isomers result in irregular shapes. We observed similar results for evaporation of DBP onto Au(111) kept at room temperature, as well as after annealing the DBP/Au(111) sample to moderate temperatures of 100 to 175 °C (Fig. S1).

The situation gradually starts to change after annealing the sample to around 200 °C. Fig. 2d is an AFM image highlighting the different kinds of molecules that were typically found after annealing the sample to a temperature of 200 °C (see Fig. S1d for an overview image). The image shows an intact DBP molecule (blue arrow) coexisting with molecules presenting either one or two lost bromines. The debrominated sides of molecules appear comparatively darker in the AFM image, suggesting a lower adsorption height which we assign to the stronger interaction of the radical with the Au(111) surface. Some of the (partially) debrominated molecules form chains, with the radicals of adjacent molecules pointing towards each other. The Laplace-filtered AFM image of such a chain in Fig. 2e enables a closer inspection of the bonding region between adjacent molecules. The apparent bond[31–33] between two monomers has a length of 3-4 Å, much too long for a covalent carbon-carbon bond. STM images of the same chain (Fig. S2) present a circular feature at certain bias voltages at the position of the apparent bond in the AFM images. Thus, the formation of such chains is rationally attributed to the bonding of adjacent monomer radicals to a coordinating Au atom. Whether that Au atom is an adatom or a surface atom that is partially lifted out of the surface plane cannot be deduced from our SPM data. The bonding of radical sites of two monomers to the same metal atom was predicted as a transition state for surface-assisted Ullmann coupling on the basis of density functional theory (DFT) calculations.[22,34] Contrary to Ag and Cu surfaces, on Au such metal-organic intermediates are observed only exceptionally.

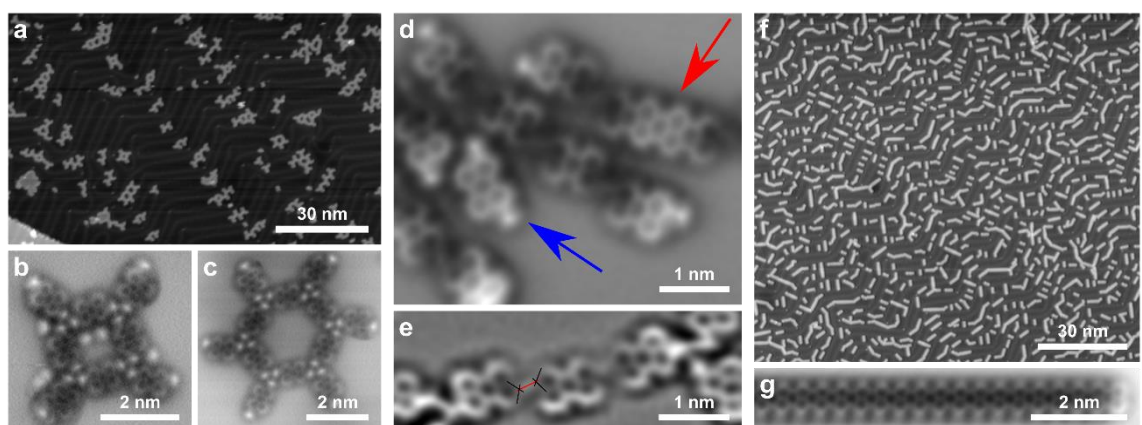

***Fig. 2*** *SPM characterization of 5-AGNR formation from DBP on Au(111). STM and AFM set points as indicated. (a) Overview STM image of the sample after deposition of DBP molecules (-1.05 V, 0.1 nA). (b, c) AFM images of self-assembled structures (b: 0.29 V, 0.05 nA, Δz = -0.4 Å; c: 0.31 V, 0.2 nA, Δz = 0.2 Å). (d) AFM image after annealing to 200 °C (-0.22 V, 0.1 nA, Δz = 0.0 Å). The blue arrow marks an intact DBP molecule and the red arrow of molecular dimer. (e) Laplace-filtered AFM image (0.2 V, 0.1 nA, Δz = 0.2 Å). The red line marks the apparent bond connecting two DBP radicals. (f) Overview STM image after annealing to 330 °C (-0.73 V, 0.1 nA). (g) AFM image of a single 5-AGNR (-0.2 V, 0.1 nA, Δz = -0.15 Å).*

The AFM image in Fig. 2d also reveals the presence of a molecule with a carbon backbone of twice the length of the DBP monomer (red arrow). Such dimer indicates that aryl-aryl homocoupling between debrominated DBP precursors along with a concomitant CDH is already possible at temperatures with partial debromination efficiency. Coexistence of both intact DBP molecules and completed intermolecular covalent coupling including CDH is highly unexpected and spans over a temperature range between 200 and 250 °C. Such behavior diverges from other GNR formation, which show well-separated temperature thresholds for polymerization upon radical formation by debromination and a subsequent completion of GNR synthesis by CDH.[4,25,35]

Despite this mixture of intermediates at medium temperatures, further annealing above 250 °C leads to the formation of well-defined, mostly straight 5-AGNRs, as demonstrated in the STM and AFM images in Figs. 2f and 2g, respectively. The overview STM image agrees with previous reports on the on-surface synthesis of 5-AGNRs from DBP on Au(111)[21,29].

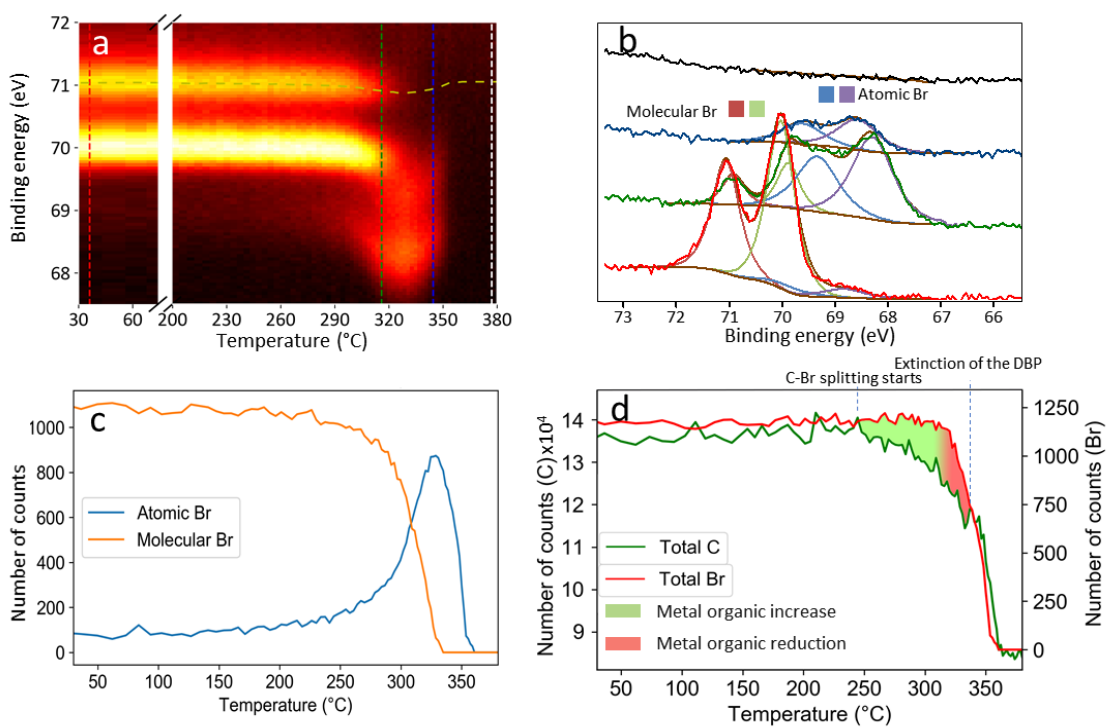

*Fig. 3.* Core level temperature dependence of DBP on Au(111) acquired at the SLS synchrotron. (a) Photoemission evolution with increasing temperature at the Br 3d energy window. Dashed yellow line represents the calculated shift in the signal caused by the change in the surface work function (explained in SI1). (b) Selected XPS spectra of the Br 3d indicated by the dashed lines in panel a. In the red spectrum there is a pair of peaks assigned to molecular Br. The blue spectrum shows the complete transformation into atomic Br and its progressive desorption from the surface, which is complete in the black spectrum. (c) Br peak intensity evolution extracted from the spectral fitting of the temperature ramp. (d) Comparison of the integrated C 1s and Br 3d signals with increasing temperature. While C desorption is linked to C-Br splitting, Br desorption is associated to the presence of H caused by cyclodehydrogenation (ribbon formation). All the XPS measurements were taken with 420 eV photon energy. For further details see "Methods".

To get precise insight into the temperature dependence of the reaction, we performed a series of synchrotron radiation fast-XPS measurements at the Pearl beamline (Swiss Light Source) while ramping up the sample temperature. A sample of ≈ 0.9 ML of DBP was prepared and placed in a resistively heated stage. At room temperature the XPS data of the Br 3d doublet (Fig. 3a and Fig. 3b (red)) relates to molecular Br. At a temperature around 240 °C this signal starts decreasing and a new doublet associated with atomic bromine shows up (Fig. 3a and Fig. 3b (green), Fig. 3c), maintaining the total Br 3d intensity unchanged (Fig. 3d). Instead, the integrated carbon signal starts decreasing along with the halogen activation onset (Fig. 3d). The carbon signal reduction is related with a partial desorption of the debrominated molecules, which are more volatile than the heavier precursors. These combined signatures mark the onset of the C-Br bond cleavage and the corresponding metal-organic intermediate formation.

The integrated Br signal including both the molecular and atomic Br contributions remains constant until a temperature around 315°C (Fig. 3d). The halogen desorption from metallic surfaces is known to be promoted by the presence of atomic hydrogen on the surface,[36,37] which causes their recombination into HBr(g) that desorbs from the surface. The cyclodehydrogenation reaction being the main experimental source of atomic hydrogen, we associate the temperature of 315 °C with the CDH onset. However, the dehalogenation of the pristine reactants is not completed until a temperature of around 340 °C (Fig. 3c). Therefore, in this temperature range ( 315°C to 340°C) there is coexistence of all reaction species (DBP, metal-organic intermediates and GNRs) on the surface. This finding matches our SPM experiments, but for a small temperature shift that is presumably due to calibration differences between the XPS and SPM setups. However, despite the absolute values, both the SPM and XPS experiments reveal a limited temperature window of a few tens of degrees were all reaction species coexist.

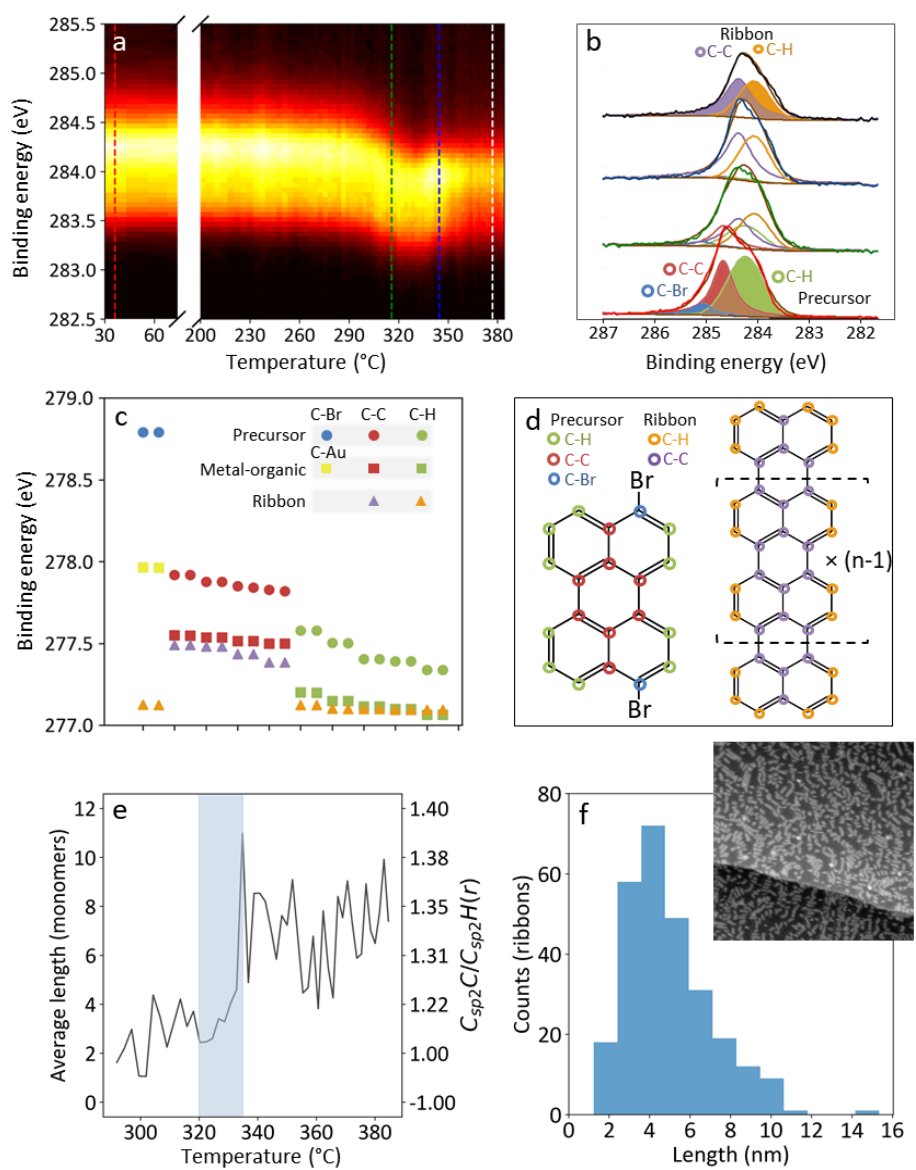

*Fig. 4*. Synchrotron XPS analysis of the annealing of DBP on Au(111) for the C1s. (a) XPS temperature evolution, where the shift in the average local workfunction caused by the Br presence has been corrected. (b) Selected XPS spectra of the C1s indicated by the dashed lines in a. (c) Core level binding energy simulations for each of the carbon atoms within the precursor molecule, metal-organic intermediates and a two-unit cell long ribbon. (d) Color-coded representation of the different categories

*into which the carbon atoms are grouped according to their core-level binding energies of c. (e) Average length of the ribbons extracted from the CC/CH intenisty ratio of the XPS spectra. The C-C/C-H signal ratio of the GNR (right scale) increases with monomer length from 1 to 1.33. (f) Average length histogram extracted from STM images recorded after the temperature ramp.*

Understanding the evolution of the temperature-dependent C1s signal (Fig. 4a) requires a more elaborate fitting procedure than the Br 3d (Fig. 4b). For this, we put forward a DFT model that provides the binding energies of each of the carbon atoms present in the reactants, metal-organic intermediates and cyclodehydrogenated ribbons (Fig. 4c), which is then used for the experimental C1s fitting (Fig. 4b). In the reactant, the calculations reveal three "groups" of carbon atoms for both racemates that properly correlate with the XPS fit components (Fig. S3). The atoms constituting each group are colored accordingly in Fig. 4d. As intuitively expected, the group with largest binding energy corresponds to C atoms bound to the more electronegative Br atoms, followed by C-C species, and C-H as the lowest binding energy (H being less electronegative). The calculated energy spread within each group is largest for the latter, which justifies the use of a substantially larger width for this component (Fig. 4b, green), while still maintaining the intensity ratio dictated by the group's stoichiometry.

The calculations for the graphene nanoribbons reveal two "groups" of carbon atoms, corresponding to the edge carbons bonded to hydrogen and those bonded only to other carbon atoms (Fig. 4d). For the fitting, the C-C/C-H ratio was left free, since it varies as the ribbons grow longer (Fig. S4). Such ratio evolution validates the adequateness of the proposed fitting and simultaneously provides complementary information about the GNR's growth dynamics, as shown later.

For the metal-organic intermediates DFT finds binding energies at values between those of the precursor and the ribbons (Fig. 4c). Such energy positions enormously complicates the experimental fittings since multiple reaction species coexist on the surface at these temperatures (Fig. S5), each contributing with a variety of carbon signals and all squeezed in the spectra within less than 2 eV. Moreover, the low concentration of Au-coordinated carbon atoms, the small temperature window of the metal-organic metastability and the fact that there are no previous XPS measurements to compare with, did not allow us to extract any rigorous conclusion about the metal-organic core levels. In essence, we avoid overfitting and refrain from blind guessing this low concentration species in the XPS data (Fig. S6/S10).

Contrarily, we can trust the fitting results from the components associated to pristine reactants and components associated to the GNRs, both of which can be nicely fitted independently of one another at the low and high temperature ends of Fig. 4a (exemplified with the red and black spectra in Fig. 4b). For the temperature-dependent data, the C1s components have been additionally shifted in energy according to the workfunction variations caused by the varying concentration of atomic Br on the surface. The applied shift is represented in Fig. 3a with the yellow, dashed line and the protocol is described in Fig. S10.

The C-C/C-H ratio in the XPS fits of the GNRs allows us to extract the average ribbon length for the relevant temperature window (Fig. 4e). It reveals three different phases in the GNR growth. In a first stage, when the reaction is still fed by the product of the dehalogenation, the GNRs maintain an approximately constant average length of around 3 monomeric units. Accordingly, the reaction does not proceed like a nucleation mechanism, where the initially formed GNRs are more likely to react. Instead, the reaction is likely to happen between single dehalogenated units that encounter on the surface. We rationalize this scenario by a larger mobility of the monomeric or dimeric units, which are thus more likely to encounter each other and react than the already formed longer ribbons.

A second phase sets in once all DBP has been fully dehalogenated, whereby the source of metal-organic units and Br is exhausted, and short active chains start decreasing in concentration. The average length of the GNRs suddenly grows over a relatively small temperature window for two reasons: (i) the steric hindrance of the Br on the surface is reduced by desorption, accelerating the reaction and (ii) the lighter short GNRs/MO monomeric and dimeric species desorb from the surface (as supported by the still decreasing C signal (Fig. 3d)), leaving room for the formation of longer GNRs.

In the last phase, the average GNR length remains again relatively constant since the more volatile GNRs have already fused or desorbed, as well as the atomic Br. There are no residual MO structures, and the only possible reaction is the uncontrolled dehydrogenation and fusion of mature GNRs, whose yield is very low at the maximum probed temperatures, justifying the absence of changes in the XPS spectra.

The resulting sample after the XPS measurements was characterized by STM at the end of the experiment (see inset of Fig. 4f), allowing for the determination of the statistical GNR length distribution after reaching the highest temperatures. The average GNR length extracted from the XPS analysis is around 7 monomers (Fig. 4e). To be comparable with the STM histogram it needs to be appropriately weighted, since the signal measured by XPS is directly proportional to the number of atoms, so that longer ribbons provide a larger signal contribution than shorter ones. Taking this into consideration, we find an excellent match in the average GNR length between both techniques, validating the previous interpretation of the growth dynamics.

At this point we draw attention on the notable and rather unexpected tendency of the studied reactants to fuse in a collinear manner that results in straight 5-AGNRs (Fig. 1e) rather than at an angle (Fig. 1f).[21,29] This is particularly noteworthy given the mixture of both *cis* (Fig. 1a) and *trans* (Fig. 1b) reactants, as well as the coexistence of parallel (Fig. 1c) and staggered (Fig. 1d) alignments in the metal-organic intermediates (Fig. 2e). The prevalence for the linear bonding motif and straight 5-AGNR formation may be expected from thermodynamic arguments. It involves the formation of a new 6-membered ring after the CDH step, in contrast to a new 5-membered ring for the kinked junction. The thermodynamic preference for the former relates to the lower strain of the structure involving exclusively hexagonal rings, as well as to the tendency to avoid spin-frustration caused by odd-membered rings in carbon $sp^2$-bonded carbon networks,[38] which reduces the stability of such structures and is typically mirrored in lower HOMO-LUMO gap values.[38,39] Indeed, we have confirmed this by DFT calculations of molecular dimers, which reveal their linear coupling product to be 0.46 eV more stable in the Gibbs free energy than the kinked one.

Nevertheless, the regioselectivity towards the benzenoid product is univocally related with the halogen activation of the precursor. In its absence, there is a clear prevalence for the kinked junctions instead, as observed in the work of Zeying Cai et al.[40] that studies the fusion of non-halogenated quaterrylene (four unit cell long 5-AGNR). This scenario has been rationalized as an effect of a kinetically driven reaction.[40] We hypothesize that, in the absence of radicals, the intermolecular coupling proceeds preferentially via an initial covalent bonding of the two molecules through the more reactive zigzag edge carbon atoms, but in a staggered configuration to minimize the steric hindrance. Thereafter, a CDH reaction in adjacent C sites will necessarily form a 5-membered ring and a kinked junction (Fig. S11a). Likewise, starting from the halogenated reactants, if a single covalent bond were directly formed after dehalogenation or after the formation of a metal-organic complex, following a conventional Ullmann coupling scenario, the molecules binding with a staggered alignment would share the structure with the intermediate assumed for the quaterrylene case. Under these circumstances, they would continue the reaction towards the kinked product (Fig. 5b). Instead, we propose that the bonding mechanism at work here derives from a concerted bond formation at the two zigzag sites by activation of only two C-H bonds facilitated by the radicals generated at the dehalogenation sites (Fig. 5a). This mechanism leads to the thermodynamically favored product featuring straight chains fully composed by 6-membered rings. Note that a similar mechanism with fully hydrogenated ends (as is the case of quaterrylene) would instead require the simultaneous abstraction of four hydrogens per fusion, which is no longer favored (Fig. S11b).[41]

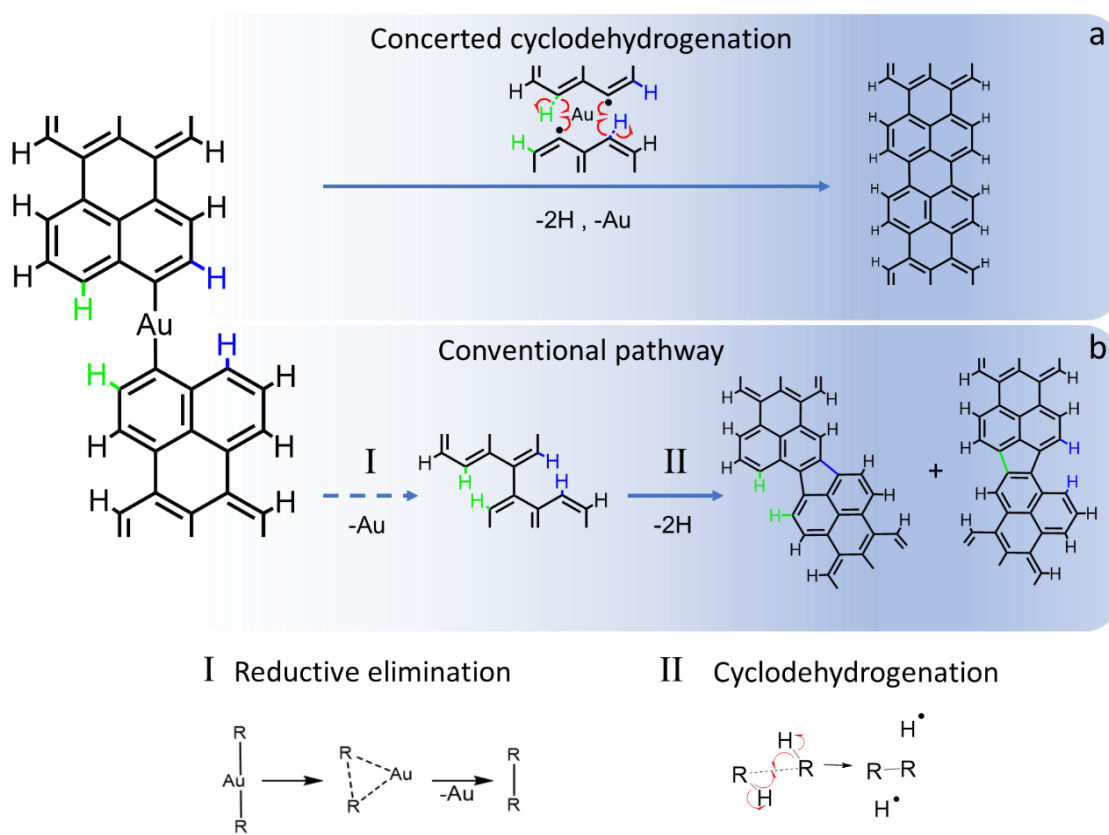

*Fig. 5*. Proposed reaction mechanisms for the prevalently formed straight (a) and the kinked junction (b).

Notably, starting from our DBP precursors the steps involved in a conventional Ullmann coupling reaction always end with kinked GNRs. We experimentally find that the prevalent straight ribbons obtained from DBP are the product of a concerted cyclodehydrogenation guided by radicals. The unfavored Ullmann coupling with DBP molecules contrasts with the reaction mechanism followed by the majority of other GNR precursors. We associate it with the steric hindrance between DBP molecules, which is promoted by their flat structure. This forces the molecules to be coplanar on the surface, leaving no space to avoid steric hindrance, as will be the case with a more flexible three-dimensional assembly.

Our proposed mechanism is closely related to the growth process found on Cu(111),[35] where the observed metal-organic intermediate associated with the molecular diradical is followed by a second metal-organic intermediate associated to tetraradical molecules that form upon activation of the remaining zigzag-site hydrogen atoms. Thereafter, the reaction proceeds toward the straight GNR junction. On Au(111), where metal-organic intermediates are known to be energetically less favorable,[24] the second metal-organic structure associated to the tetraradicals was not detected, and the reaction proceeds directly to the concerted bond formation. However, on Cu(111) no co-existence of the different intermediates was observed, and the resulting AGNRs were shorter and often kinked.[35] This is in stark contrast to the results for Au(111) presented here, where despite the temporary presence of a 'messy' mixture of molecules due to the coexistence of all reaction species, primarily long, straight AGNRs are formed.

## Conclusions:

We provide detailed insight into the growth process of 5-AGNRs from dibromoperylene through a multi-technique approach. Scanning probe microscopy experiments revealed that for a specific temperature window there is a simultaneous presence of pristine reactants, metal-organic intermediates and fully cyclodehydrogenated GNRs. A detailed analysis of the carbon and bromine core-level intensity evolution from complementary temperature-dependent XPS experiments provides insight into the GNR growth dynamics and average length. Our experimental results supported with DFT allow us to unambiguously show that the reaction path starting from these planar precursor molecules to ultimately form 5-AGNRs does not follow the expected sequence of Ullmann coupling and subsequent cyclodehydrogenation. Here, a concerted mechanism is proposed instead, involving the radical-guided activation of C-H bonds in both reactants, to simultaneously form two C-C bonds that result in a benzenoid ring and a straight intermolecular junction. The insight into this new reaction mechanism may help understanding the chemical processes followed by other reactants,[42] as well as predicting the product structures from other rigid and planar halogenated precursors.

## Methods:

SPM measurements: SPM characterization of the sample was carried out with a Createc LT-STM/AFM system under ultra-high vacuum conditions (base pressure ~$10^{-10}$ mbar). As substrate we used a Au(111) single crystal, which was cleaned by standard sputtering and annealing cycles. DBP was thermally sublimed onto the clean sample held either at or slightly below room temperature. The sample temperature during annealing was monitored with a pyrometer. After preparation, the sample was inserted into the STM/AFM housed within the same vacuum system and operated at a temperature of 5 K. The scanner was equipped with a qPlus force sensor [43] with resonance frequency $f_0$ of ~30.7 kHz, quality factor $Q$ of ~100k, and spring constant $k$ of ~1.8 kN/m. STM measurements were performed in the constant-current mode with the bias voltage applied to the sample. AFM measurements were performed in frequency modulation mode [44] at constant height, with a tip oscillation amplitude of 50 pm and at 0 V bias voltage. Positive (negative) tip height offsets *Δz* correspond to a decrease (increase) of the tip-sample distance with respect to the STM set point. We used CO-functionalized tips for bond-resolved AFM imaging.[30] CO was dosed onto the sample through a leak valve and picked up from the surface with the SPM tip as described previously.[45] It should be noted that at the elevated temperatures at which the intermediates are formed, the molecules are very mobile and can explore a wider region of the potential energy surface, whereas the SPM analysis can only identify the lowest-energy structure of the reaction intermediates upon cooling down to SPM imaging temperatures (5 K in our case).

XPS measuremens: The soft X-ray photoelectron spectroscopy (XPS) measurements were performed with a photon energy of 420 eV at the Pearl Light Beam in the Swiss Light Source, at Villigen (Switzerland). Specifications about the instrumentation are detailed in the technical documentation of the beamline[46]. The heating rate during the temperature ramp was approximately 0.7 K/minute.

Theoretical calculations: Theoretical calculations were performed in the frame of the density functional theory (DFT) using the B3LYP hybrid functional.[47,48] Geometry optimizations were performed with the cc-pVDZ basis set, and all the stationary points were confirmed to be minima of the potential energy surface through a second-derivative analysis. Single point calculations with the cc-pVTZ basis set were performed to get the energies of the core orbitals. Hybrid functionals with large basis sets provide valuable information for XPS.[49,50] All the calculations were done using the Gaussian16 software.[51]

Temperature considerations: Two pyrometers in different geometries were used to measure the sample temperature in the STM and XPS systems. For each separate system the experimental observations for a given annealing temperature were very reproducible. However, obtaining identical temperatures between the two different experimental set-ups is generally impossible without adding correction factors because of differences in pyrometer calibration and influence of stray light on the temperature measurement.

## Acknowledgements:

We acknowledge financial support from the Spanish Agencia Estatal de Investigación (AEI) and the European Regional Development Fund (FEDER) (Grants No. PID2019-107338RB-C63 and PID2019-107338RB-C64) and the Academy of Finland (project numbers 311012 and 314882). We acknowledge the Paul Scherrer Institut, Villingen, Switzerland for provision of synchrotron radiation beamtime at PEARL Beamline of the SLS and would like to thank N. P. M. Bachellier for assistance. We thank R. Fasel and R. Widmer for provision of substrate and sample holder for the synchrotron experiments.

## Supplementary Information:

# Order from a mess: the growth of 5-armchair graphene nanoribbons

Alejandro Berdonces-Layunta,[1,2,†] Fabian Schulz,[3,4,†,*] Fernando Aguilar-Galindo,[1,†] James Lawrence,[1,2] Mohammed S. G. Mohammed,[1,2] Jorge Lobo-Checa,[5,6] Peter Liljeroth,[3] Dimas G. de Oteyza.[1,2,7,*]

[1] Donostia International Physics Center, 20018 San Sebastián, Spain

[2] Centro de Física de Materiales, 20018 San Sebastián, Spain

[3] Department of Applied Physics, Aalto University, FI-00076 Aalto, Finland

[4] Fritz Haber Institute of the Max Planck Society, 14195 Berlin, Germany

[5] Instituto de Nanociencia y Materiales de Aragón, 50009 Zaragoza, Spain

[6] Dpto. de Física de la Materia Condensada, Universidad de Zaragoza, 50009 Zaragoza, Spain

[7] Ikerbasque, Basque Foundation for Science, Bilbao, Spain

† these authors contributed equally

* correspondence to: schulz@fhi-berlin.mpg.de, d_g_oteyza@ehu.es

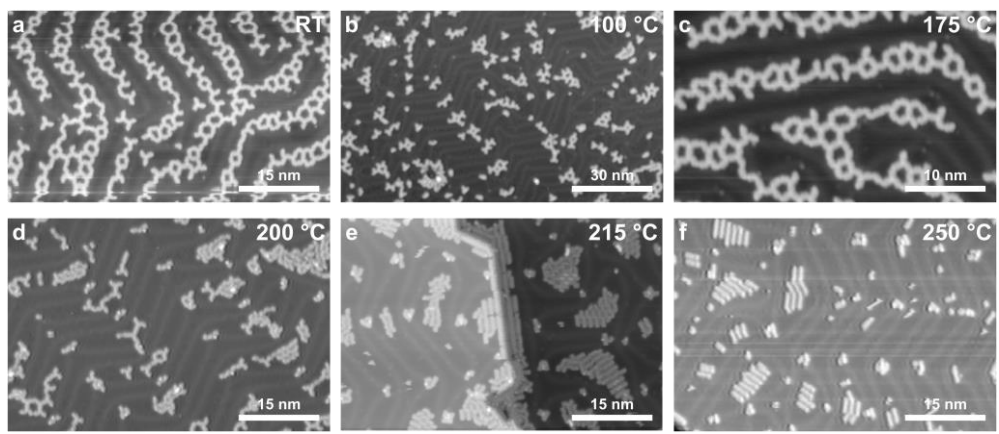

**Fig. S1.** STM overview images of the DBP/Au(111) sample after annealing to various temperatures. STM set points as indicated. (a) High coverage of DBP on Au(111) as-deposited at room temperature (-0.53 V, 0.11 nA). (b) The sample of Fig. 2a in the main manuscript (low coverage of DBP deposited at 250 K) after annealing to 100 °C (0.30 V, 0.20 nA). (c) The sample of Fig. S1a after annealing to 175 °C (-0.78 V, 0.12 nA). (d) After annealing to 200 °C (0.48 V, 0.33 nA). (e) After annealing to 215 °C (-0.95 V, 0.05 nA). (f) After annealing to 250 °C (-0.58 V, 0.07 nA).

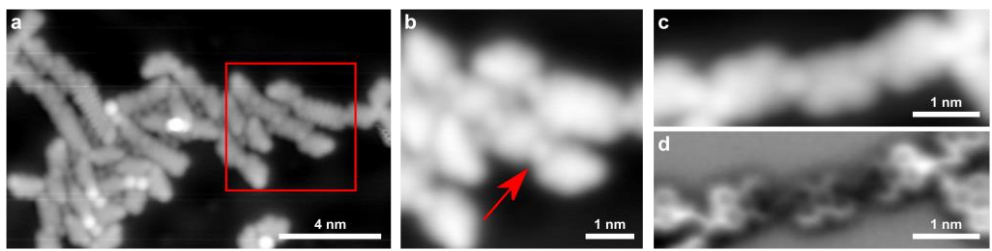

**Fig. S2.** Coexistence of all reaction species. STM and AFM set points as indicated. (a) STM image of a typical molecular aggregate observed after annealing DBP/Au(111) to 200 °C (-0.22 V, 0.08 nA). (b) Corresponding STM image (region marked with a red square in panel a) of the AFM image in Fig. 2d in the main manuscript (1.31 V, 0.10 nA). The red arrow marks a circular feature linking two (partially) debrominated DBP monomers. (c, d) Corresponding STM and AFM image of the Laplace-filtered AFM image in Fig. 2e, respectively (0.20 V, 0.10 nA, $\Delta z$ = 0.2 Å).

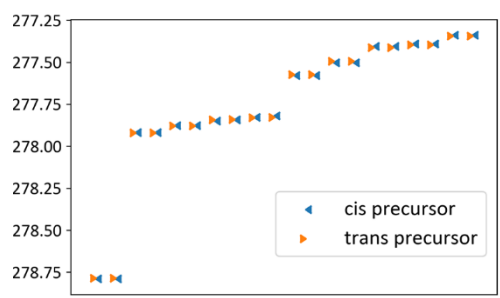

**Fig. S3.** C1s XPS simulations for the two racemates of the precursor.

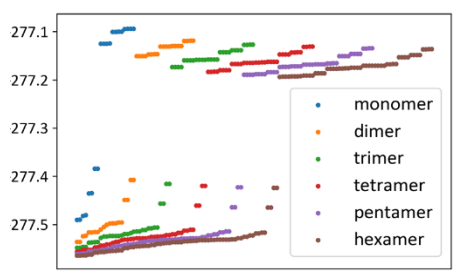

**Fig S4.** C1s XPS simulations for different ribbon lengths. Note how the carbons energetically split in two separate groups: C-C and C-H. The four signals around 277.4 eV that stand out belong to the carbons of the zigzag edge.

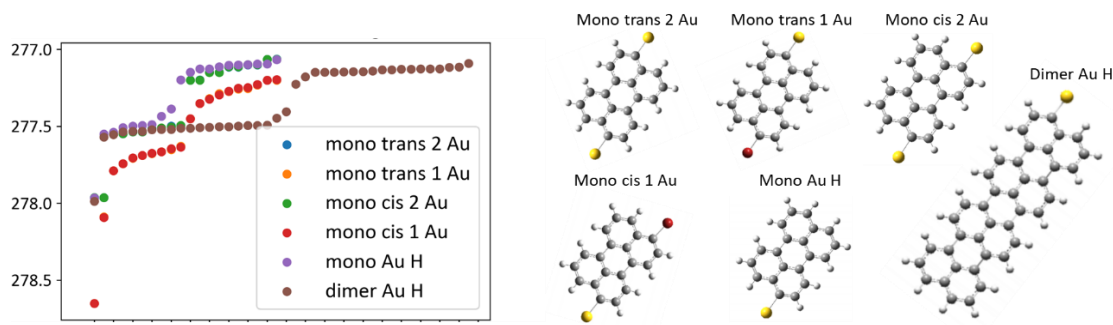

**Fig S5.** C1s XPS simulations for the different types of MO observed in the AFM images.

### Supplementary note 1. C1s XPS: analysis and interpretation

The starting considerations include our experimental observation of a metal-organic intermediate within a limited temperature range. However, as readily extracted from our SPM experiments, even within the right temperature window the number of molecules forming the metal-organic intermediate is limited, coexisting with the other molecular species. Since only two atoms per precursor molecule coordinate at most to metal atoms, the stoichiometric ratio of these carbon species is extremely limited throughout the whole growth process.

Below we describe the different approaches that have been followed to fit the C1s spectra.

*First approach: Precursor and ribbons*.

As a starting point, we have fitted the spectra associated to as-deposited molecules and to the GNRs, as extracted from the first and last spectra during our temperature ramp.

For the as-deposited molecules the carbon signal was modelled with three components based on theoretical core level energy calculations. The carbons bonded to hydrogen (C-H) are the lowest in energy. Then, the fully substituted inner Carbons (C-C). The highest binding energy is displayed by those carbons attached to Br (C-Br). The ratio between the components was fixed according to the molecule's stoichiometry (C-H)/(C-C)/(C-Br) = 10/8/2. Parameters like shape, Full-Width Half Minimum (FWHM), relative and absolute position of the peaks were optimized to get the best possible fit. The resulting fit provides notable differences in the FWHM of each of the components, which are in qualitative agreement with the energy spread theoretically predicted for the core levels within each group (see Fig. 4c in main text).

For the GNRs the calculations predict two different groups of carbons, one for the external C-H carbons and one for the internal ones. In this case, the ratio C-C/C-H depends on the average length of the fully grown GNRs according to $ratio = \frac{12N-4}{8N+4}$; where $N$ is the number of precursor units.

As a result, we can estimate the average ribbon length from the fitted C-C/C-H ratio.

In spite of the highly asymmetric line shape of the spectra, each of the fitting components is maintained symmetric. Although asymmetric peaks have been used in the study of GNRs earlier, we believe that their use is not really justified here, because the bandgap of these GNRs prevents the low energy excitations that are normally imposed to justify the high binding energy "tails" of the core level peaks.

Within our first approach, the fitting of all spectra consisted in a linear combination of the components associated to reactants and to GNRs. In Fig. S6 we include sample spectra for the two reference situations, along with a list of the free and fixed parameters that the system could modify.

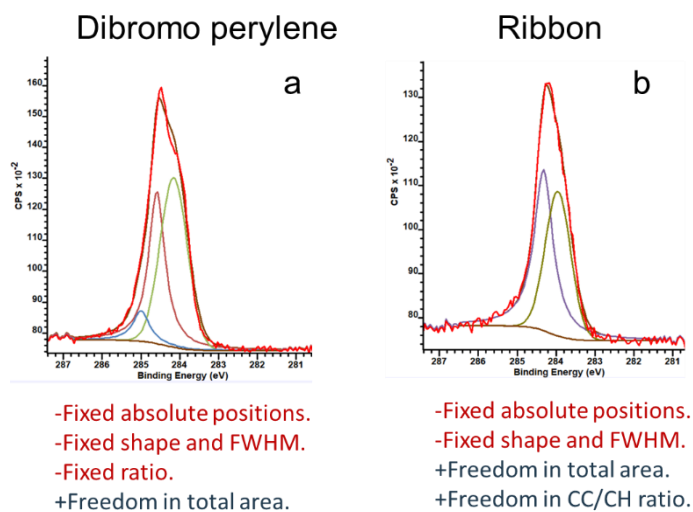

**Fig. S6.** Sample spectra for the two reference situations, namely as-deposited molecules (a) and GNRs (b), along with the parameters that have been fixed and fitted for the various components, respectively.

With these three degrees of freedom, the resulting fit seemed to follow the expected tendency. However, two main problems arise:

-The temperature window of active reactivity is poorly fitted, in particular at the low binding energy side (See Fig. S7).

-The ratio between the peaks of the ribbon exceeds the limits (C-C/C-H) = 0.66 ≤ r ≤1.5.

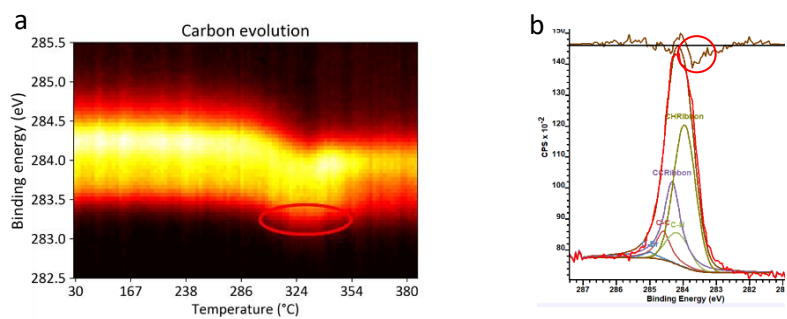

**Fig. S7**. The first fitting approach failed in addressing the XPS signal under 284 eV. It caused a systematic error and an unphysical C-C/C-H ratio.

*Second approach: Addition of the metal-organic.*

To overcome the inconsistencies obtained from the previous fitting approach, we tried adding the metal-organic signal to fill the gaps left by the other two compounds. The signal associated to DBP and ribbons, as fitted in the first and last spectra, is maintained. For the metal-organic, we employed the less reliably fit spectra from the previous approach, which was assumed to contain the largest metal-organic contribution.

The relative energies for the C-C/C-H components of the metal-organic molecules have been maintained as in the previous cases. A lower binding energy for a metal-organic may be justified by a significant inductive effect caused by the Au atom. To be consequent, we placed the coordinated carbon peak (C-Au) at lower binding energies, as the literature conventionally does for the other coinage metals (Ag, Cu), although contradicting the core level energies obtained from our DFT calculations (Fig. 4c).

Representative sample spectra of the three different scenarios during the temperature ramp are displayed in Fig. S8, along with the parameters that are fitted or fixed, respectively.

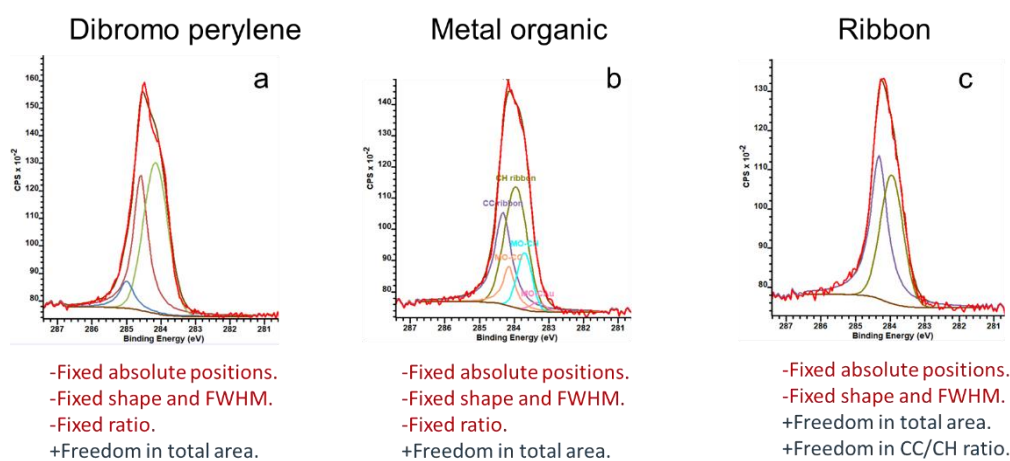

**Fig. S8**. Degrees of freedom and conditions of the new fitting.

With this new variable, the fitting improves, but the temperature-dependent amount of each component (Fig. S9) shows an unreasonable evolution with the (metal-organic) intermediate, since it maximizes with the (GNR) final product, instead of before. We therefore also discarded these fitting results.

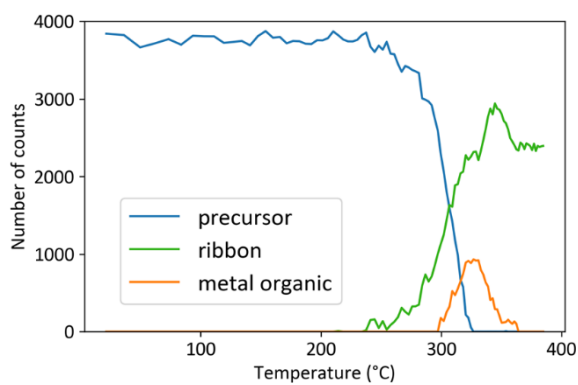

**Fig. S9** Temperature-dependent evolution of the relative amounts of as-deposited, metal-organic and GNR molecules, as obtained within this second fitting approach including the metal-organic species.

*Third approach: Addition of the Br-associated work function modulation.*

The discrepancies on the low binding energy side obtained with our first fitting approach (Fig. S6) may also be an effect of the atomic Br on the local work function. In a new approach, we proceed fitting with a linear combination of as-deposited molecules and GNRs, in combination with a work function modulation that correlates with the concentration of atomic Br on the metal surface. Such optimal fitting result corrected the inconsistencies or deficiencies from the previous two approaches.

We estimate the shift in the local work function analyzing the DBP Br signal, which displays the following advantages: it is associated with unchanged molecules and can therefore be associated to work function modifications, it can be followed univocally along the reaction and, being bound to the molecular carbon backbone, the effect on the carbon atoms can be assumed to be similar. Figure S10a shows the rigid shift of the DBP Br doublet during the temperature ramp, whereas Fig. S10b shows its linear correlation with the concentration of atomic Br on the surface. With this correlation at hand, we can estimate the associated shift of the carbon core levels throughout the temperature ramp, as plotted in Fig. S10c, and apply it to the fitting of the temperature-dependent spectra.

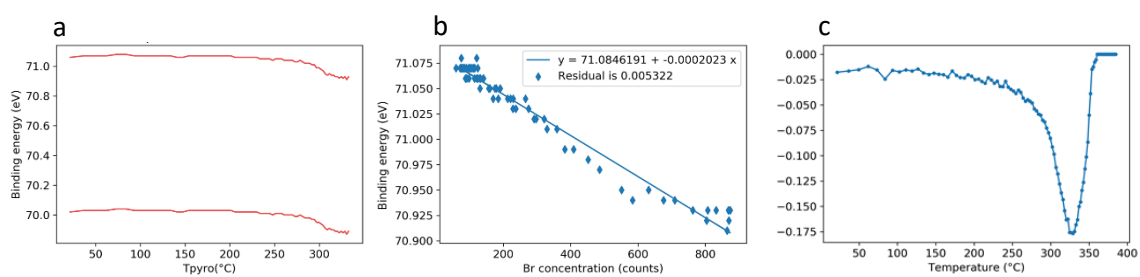

**Fig. S10**. A) Position of the pairs of bromine peaks with the temperature. Although the atomic Br peak shifts dramatically, the effect is also observed in the molecular Br. It has been proposed previously that the presence of atomic Br in the surface can change the average work function. B) Linear correlation between the Br concentration and the binding energy shift of the molecular Br. C) Estimated shift of the carbon XPS, as extracted from applying the function of B) to the atomic Br concentration.

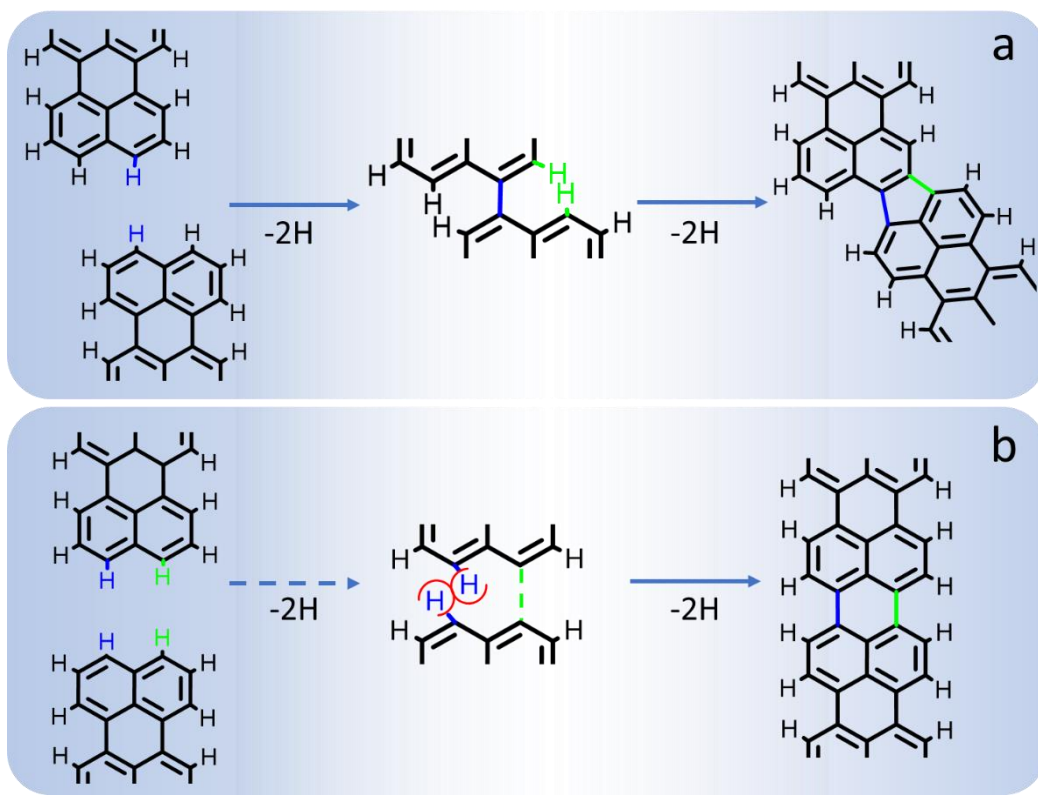

**Fig. S11.** Proposed mechanism for the halogen-free molecular coupling. There is an expected two-step mediated reaction. With submonolayer coverages, the reaction is regioselective towards their kinked fusion (a), kinetically determined due to the steric hindrance in the straight reaction pathway (b).